\documentclass[11pt]{article}
\usepackage{amssymb}
\usepackage{amsmath}
\usepackage{amscd}
\usepackage{latexsym}

\catcode`@=11
\renewcommand{\section}{\@startsection{section}{1}{0pt}{\medskipamount}
{\medskipamount}{\large\bf}}
\numberwithin{equation}{section}
\catcode`@=12
\def\a{\alpha}
\def\b{\beta}

\def\de{\delta}
\def\eps{\epsilon}

\def\vk{\varkappa}

\def\th{\theta}

\newcommand{\unity}{{\bf{1}}}
\newcommand{\Cbb}{\mathbb C}
\newcommand{\R}{\mathbb R}

\newcommand{\Gcal}{{\cal G}}
\newcommand{\Acal}{{\cal A}}

\newcommand{\Hcal}{{\cal H}}
\newcommand{\Ical}{{\cal I}}
\newcommand{\Mcal}{{\cal M}}
\newcommand{\Fcal}{{\cal F}}

\newcommand{\Ncal}{{\cal N}}

\newcommand{\gfrak}{{\mathfrak g}}
\newcommand{\mfrak}{{\mathfrak m}}
\newcommand{\hfrak}{{\mathfrak h}}

\newcommand{\bari}{{\bar{\imath}}}
\newcommand{\hati}{{\hat{\imath}}}
\newcommand{\barj}{{\bar{\jmath}}}
\newcommand{\hatj}{{\hat{\jmath}}}

\def\tr{\textrm{tr}}
\def\diff{\textrm{d}}
\def\pa{\mbox{$\partial$}}
\def\sfrac#1#2{{\textstyle\frac{#1}{#2}}}
\def\+{\dagger}
\def\={\ =\ }
\def\und{\qquad\textrm{and}\qquad}
\def\and{\quad\textrm{and}\quad}
\def\with{\quad\textrm{with}\quad}
\def\for{\quad\textrm{for}\quad}

\def\Gr{\mathrm{Gr}}

\begin{document}

\begin{titlepage}
\setcounter{page}{0}

\hspace{2.0cm}

\begin{center}

{\LARGE\bf
Skyrme and Faddeev models in the low-energy limit\\[3mm]
 of 4d Yang-Mills-Higgs theories
}

\vspace{12mm}

{\Large
Olaf Lechtenfeld${}^{\+\times}$ \ and \  Alexander D. Popov${}^\+$
}\\[10mm]
\noindent ${}^\+${\em
Institut f\"ur Theoretische Physik,
Leibniz Universit\"at Hannover \\
Appelstra\ss{}e 2, 30167 Hannover, Germany
}\\
{Email: alexander.popov@itp.uni-hannover.de}
\\[5mm]
\noindent ${}^\times${\em
Riemann Center for Geometry and Physics,
Leibniz Universit\"at Hannover \\
Appelstra\ss{}e 2, 30167 Hannover, Germany
}\\
{Email: olaf.lechtenfeld@itp.uni-hannover.de}

\vspace{20mm}

\begin{abstract}
\noindent 
Firstly, we consider U($N_c$) Yang--Mills gauge theory on $\R^{3,1}$ with $N_f > N_c$ flavours
of scalar fields in the fundamental representation of U($N_c$). The moduli space of vacua
is the Grassmannian manifold $\Gr(N_c,N_f)$. It is shown that 
for strong gauge coupling this 4d Yang--Mills--Higgs theory reduces to the Faddeev 
sigma model on $\R^{3,1}$ with $\Gr(N_c,N_f)$ as target.
Its action contains the standard two-derivative sigma-model term as well as the four-derivative Skyrme-type term,
which stabilizes solutions against scaling. Secondly, we consider a Yang--Mills--Higgs model with $N_f\,{=}\,2N_c$
and a Higgs potential breaking the flavour group U$(N_f)\,{=}\,$U($2N_c$) to U$_+(N_c){\times}$U$_-(N_c$), 
realizing the simplest $A_2\oplus A_2$-type quiver gauge theory. The vacuum moduli space of this model 
is the group manifold U$_h(N_c)$ which is the quotient of U$_+(N_c){\times}$U$_-(N_c$) by its diagonal subgroup. 
When the gauge coupling constant is large, this 4d Yang--Mills--Higgs model reduces to the Skyrme sigma model 
on $\R^{3,1}$ with U$_h(N_c)$ as target. Thus, both the Skyrme and the Faddeev model arise as effective 
field theories in the infrared of Yang--Mills--Higgs models.

\end{abstract}

\end{center}
\end{titlepage}

\section {Introduction and summary}

\noindent In 1975, Faddeev introduced a (3+1)-dimensional SU(2)/U(1) coset sigma model that includes a term
quartic in derivatives to stabilize classical solutions~\cite{Fa}. This model is similar to the Skyrme model~\cite{Sk}, 
which features maps from $\R^{3,1}$ into SU(2). Despite their similarity, these models are quite different from one another. 
Topological solitons of the Skyrme model have a point-like core and are supposed to describe baryons and nuclei 
(see e.g.~\cite{ZB} for a review and \cite{BMSW, GHKMS, AHRW} for some recent works). 
On the other hand, solitons in the Faddeev model take the form of stable knotted strings characterized by the Hopf charge 
(homotopy class of maps $S^3\to S^2$). It is conjectured that Faddeev-model solitons describe glueballs 
(see e.g.~\cite{Fad}-\cite{Ha1} for reviews).

\medskip

The standard Skyrme model~\cite{Sk} supposedly describes pions.
Other mesons can be incorporated into an {\it extended\/} 4d Skyrme model, which is obtained from 5d Yang--Mills theory on an 
AdS-type manifold $M^5$ with boundary $\pa M^5=\R^{3,1}$ as derived from 
D-brane configurations in string theory and the holographic approach~\cite{SS} (see e.g.~\cite{EEKT, KMS, Sut2} for reviews).
This extended Skyrme model also arises in the adiabatic limit of the 5d Yang--Mills system on $\R^{3,1}\times\Ical$, where
$\Ical$ is a short interval~\cite{ILP1}.\footnote{
The adiabatic approach was used in field theory for the first time by Manton~\cite{Ma}. 
For a review of this approach see~\cite{MS, WY}; brief discussions can be found e.g.~in~\cite{HS}-\cite{ILP2}.}
Similarly, also an {\it extended\/} 4d Faddeev model can emerge in a low-energy limit of 
5d maximally supersymmetric Yang--Mills theory with its five adjoint scalars~\cite{LePo}. In contrast to the extended 
Skyrme model, for the extended Faddeev model one needs to keep one of the five adjoint scalars and must modify the fifth 
dimension from $\Ical$ to the half-line $\R_+$. The boundary conditions required for the reduction to $\R^{3,1}$
are encoded in Nahm equations along the fifth dimension~\cite{GMT, Assel}, which reduce to a ``baby'' Nahm equation on $\R_+$
for one adjoint scalar~\cite{LePo}. 
 
\medskip

Quantum chromodynamics (QCD) as well as Yang--Mills theory are strongly coupled in the infrared limit, and hence the perturbative 
expansion for them breaks down. In the absence of a quantitative understanding of non-perturbative QCD, convenient alternatives
at low energy are provided by effective models among which nonlinear sigma models play an important role, especially the Skyrme and
Faddeev models. Both models are the standard two-derivative sigma models on  $\R^{3,1}$ with a compact Lie group $G$ and a coset 
space $G/H$ as target spaces, respectively, completed with a four-derivative term which stabilizes classical solutions against scaling.
In the Faddeev model, $H$ is a closed subgroup of $G$ such that $G/H$ is a coadjoint orbit.

\medskip

As we discussed above, both Skyrme and Faddeev models can be obtained as low-energy limit of 5d Yang--Mills--Higgs (YMH) theories on the 
classical level. On the other hand, in the strong-coupling or infrared limit, many YMH models on $\R^{d-1, 1}$ with $d\ge 2$ reduce 
to standard two-derivative sigma models governing maps from $\R^{d-1,1}$ to a moduli space of Higgs vacua. In other words, 
YMH theories flow in the infrared to sigma models on the same space $\R^{d-1,1}$ (see e.g.~\cite{Eto} and references therein). 
For YMH models which are bosonic parts of supersymmetric QCD in $d{=}4$, these classical moduli spaces are non-trivial K\"ahler or 
hyper-K\"ahler manifolds~\cite{Sei, BW, Eto}. 
Here we will show that the four-derivative Skyrme term also naturally appears in these four-dimensional YMH models in the framework 
of the adiabatic approach.\footnote{
Some steps in the derivation of Skyrme terms from YMH models were taken in~\cite{Bab, Moh}, but for a different class of YMH models
and without using the adiabatic method.}
To summarize, we demonstrate that both the Skyrme model and the Faddeev model occupy an infrared corner of 4d YMH models 
related with $\Ncal{=}\,2$ supersymmetric QCD.

\bigskip

\section {Yang--Mills--Higgs model}

\noindent {\bf Notation.} 
On Minkowski space $\R^{3,1}\ni x^\mu$ with the metric $(\eta_{\mu\nu})={\rm diag} (-1,1,1,1)$, 
we consider U($N_c$) gauge theory with $N_f$ flavours of scalar fields in the fundamental representation of U($N_c$), 
combined in an $N_c\times N_f$ matrix $\Phi$. A gauge potential $\Acal=\Acal_{\mu}\,\diff x^\mu$ and the Yang--Mills field 
$\Fcal=\diff\Acal+\Acal\wedge\Acal$ take values in the Lie algebra $u(N_c)$. Its components read 
$\Fcal_{\mu\nu}=\pa_\mu\Acal_\nu-\pa_\nu\Acal_\mu+[\Acal_\mu,\Acal_\nu]$, where $\pa_\mu:=\pa /\pa x^\mu$ and $\mu,\nu=0,1,2,3$. 
For the generators $I_\hati$ of the gauge group~U($N_c$) we use the standard normalization $\tr(I_\hati I_\hatj)=-\sfrac12\,\de_{\hati\hatj}$. 

\medskip

\noindent {\bf Transformations of fields.} 
The covariant derivative of the complex Higgs field $\Phi$ in the bi-fundamental representation of U($N_c$)$\times$U($N_f$) with $N_f>N_c$ reads
\begin{equation}\label{2.1}
D_\mu\Phi \= \pa_\mu\Phi + \Acal_\mu\Phi
\end{equation}
since the U($N_f$) flavour group acts on $\Phi$ only by global transformations
\begin{equation}\label{2.2}
\Phi\mapsto\Phi\,g^{}_f\ .
\end{equation}
We denote by $\Gcal$ the infinite-dimensional group $C^\infty (\R^{3,1},$ U($N_c$)) of gauge transformations which are parametrized 
by $g_c(x)\in\Gcal$ for $x\in \R^{3,1}$. Then $\Acal$ and $\Phi$ are transformed as
\begin{equation}\label{2.3}
\Acal\ \mapsto\ {}^{g_c}\Acal\=g_c\Acal g_c^{-1}+g_c\diff g_c^{-1}
\und \Phi\ \mapsto\ {}^{g_c}\Phi\=g_c\Phi\ .
\end{equation}
For the infinitesimal action of $\Gcal$ we have
\begin{equation}\label{2.4}
\Acal\ \mapsto\ \de_{\eps_c}\Acal \=\diff\eps_c-[\Acal , \eps_c] 
\und \de_{\eps_c}\Phi\=\eps_c\Phi \ \with \ g_c=\exp(\eps_c)\ .
\end{equation}
Similarly, for the U($N_f$) flavour symmetry we have 
\begin{equation}\label{2.5}
\de_{\eps_f}\Acal \= 0\und \de_{\eps_f}\Phi\=\Phi\,\eps_f\ ,
\end{equation}
where $\eps_c\in\,$Lie$\Gcal=C^\infty (\R^{3,1}, u(N_c))$ and $\eps_f\in u(N_f)$.

\medskip

\noindent {\bf Lagrangian.} 
We consider the Yang--Mills--Higgs (YMH) action functional 
\begin{equation}\label{2.6}
S\=-\int_{\R^{3,1}} \!\!\!\!\diff^4x\ \bigl\{
\tr\,\bigl(\sfrac{1}{2e^2}\,\Fcal^\+_{\mu\nu}\Fcal^{\mu\nu} +  D_{\mu}\Phi (D^{\mu}\Phi)^\+\bigr) +\sfrac{e^2}{4}\, V(\Phi )\bigr\}\ ,
\end{equation}
where $\+$ denotes Hermitian conjugation, $e$ is the gauge coupling constant, and 
\begin{equation}\label{2.7}
V(\Phi) \= \tr\,\bigl(M^2\unity_{N_c} - \Phi\Phi^\+\bigr)^2
\end{equation}
is the Higgs potential with a mass  parameter $M$. 
The Lagrangian from (\ref{2.6}) is related with the bosonic part of the Lagrangian for $\Ncal{=}\,2$ supersymmetric QCD,
and such Lagrangians are often considered in the literature (see e.g.~\cite{Eto, Tong} and references therein). 

The energy density $\Hcal$ of YMH configurations described by  (\ref{2.6})  is  
\begin{equation}\label{2.8}
\Hcal \= \tr\,\bigl(\sfrac{1}{e^2}\,\Fcal^\+_{0a}\Fcal_{0a}+ D_0\Phi (D_0\Phi )^\+
+ \sfrac{1}{2e^2}\,\Fcal^\+_{ab}\Fcal_{ab}+ D_a\Phi (D_a\Phi )^\+ \bigr) + \sfrac{e^2}{4}\, V(\Phi)\ ,
\end{equation}
where $a,b=1,2,3$. Here both $V(\Phi)$ and $\Hcal$ are positive-semidefinite and gauge-invariant functions. 
They are also U($N_f$)-invariant. 

\medskip

\noindent {\bf Vacua.} 
A YMH vacuum configuration $(\hat\Acal,\hat\Fcal,\hat\Phi)$ is defined by the vanishing of the energy density~(\ref{2.8}). 
This is achieved by
\begin{equation}\label{2.9}
\hat\Fcal_{\mu\nu}=0\ ,\quad \hat{D}_{\mu}\hat\Phi=0 \und V(\hat\Phi )=0\ ,
\end{equation}
where the last equation defines the Higgs vacuum manifold. Denote by $\widetilde\Mcal$ the space of solutions of (\ref{2.9}) with $\hat\Acal=0=\hat\Fcal$ and 
$\hat\Phi \in\,$Mat($N_c, N_f; \Cbb$) (complex $N_c\times N_f$ matrices) such that
\begin{equation}\label{2.10}
\hat\Phi\hat\Phi^\+=M^2\unity_{N_c}\ ,
\end{equation}
i.e.~$\widetilde\Mcal$ is the space of solutions to~(\ref{2.10}). The group U($N_c$) acts freely on $\widetilde\Mcal$ by left multiplication, 
$\hat\Phi\mapsto g_c\hat\Phi$. It is not difficult to show  \cite{KNv2} that $\widetilde\Mcal$ is fibred over the Grassmannian 
\begin{equation}\label{2.11}
\Gr(N_c, N_f)\= [{\rm U}(N_c)\times {\rm U}(N_f{-}N_c)] \backslash {\rm U}(N_f)\ =:\ \Mcal
\end{equation}
with the projection
\begin{equation}\label{2.12}
\pi\ :\quad \widetilde\Mcal\stackrel{{\rm U}(N_c)}{\longrightarrow} \Mcal
\end{equation}
and the group U($N_c$) as fibres.

It is important to distinguish between the Higgs field $\Phi$ depending on $x\in\R^{3,1}$ and the vacua $\hat\Phi\in\,$Mat($N_c, N_f; \Cbb$), which solve~(\ref{2.10}). 
The moduli space of vacua $\Mcal$ is the Grassmannian (\ref{2.11}), any element of which can be obtained from a reference vacuum $\hat\Phi_0$. 
We choose $\hat\Phi_0=(\unity_{N_c}\  {\bf 0}_{N_c\times (N_f-N_c)})$ so that the isotropy group of $\hat\Phi_0$ for the right U($N_f$) action is 
\begin{equation}\label{2.13}
{\rm U}(N_c)\times  {\rm U}(N_f{-}N_c)\=\bigl\{ g^{}_f\in{\rm U}(N_f):\ \hat\Phi_0\,g^{}_f=g_c\hat\Phi_0\ \mbox{for some}\ g_c\in{\rm U}(N_c)\bigr\}\ .
\end{equation}
It is obvious \cite{KNv2} that such $g^{}_f$ have the form diag$(g_c, g^{}_{f-c})$ with $g^{}_{f-c}\in\,$U($N_f{-}N_c$). 
In other words, the right action of the isotropy group ${\rm U}(N_c)\times  {\rm U}(N_f{-}N_c)$ on $\hat\Phi_0$ is equivalent to the left action 
of the gauge group U($N_c$), and we simply have
\begin{equation}\label{2.14}
\widetilde\Mcal\={\rm U}(N_f{-}N_c)\backslash {\rm U}(N_f)\ .
\end{equation}

\bigskip

\section{Moduli space of vacua}

\noindent {\bf Geometry of $\Gr(N_c, N_f)$.} 
The space $\widetilde\Mcal$ in (\ref{2.12}) parametrizes all vacua for the model (\ref{2.6}), \
and the Grassmannian $\Mcal$ in (\ref{2.11}) and (\ref{2.12})  parametrizes gauge inequivalent vacua, i.e.~the vacuum moduli space. 
Both $\widetilde\Mcal$ and  $\Mcal$ are homogeneous spaces with a right action of U($N_f$). 
Note that right cosets can be changed to left cosets by interchanging $\hat\Phi$ with $\hat\Phi^\+$.

Let $\mfrak$ be the tangent space to the Grassmannian $\Mcal$ at the fixed point $\hat\Phi_0$. Then we have the splitting 
\begin{equation}\label{3.1}
u(N_f) \=\mfrak \oplus u(N_c)\oplus u(N_f{-}N_c)\ ,
\end{equation}
and $\tilde \mfrak = \mfrak \oplus u(N_c)$ can be identified with the tangent space of $\widetilde\Mcal$ at any given point. 
For $u(N_f)$ we choose a basis 
\begin{equation}\label{3.2}
\{I_i\} = \{I_{\bari},I_{\hati},I_{i'}\} \ \with\ \begin{cases}
\ \bari\!\!&=\ 1,\ldots,\dim\mfrak = 2N_c(N_f{-}N_c)\ , \\
\ \hati\!\!&=\ \dim\mfrak{+}1,\ldots,\dim\mfrak{+}N_c^2\ ,\\
\ i'   \!\!&=\ \dim\mfrak{+}N_c^2+1,\ldots,\dim\mfrak{+}N_c^2{+}(N_f{-}N_c)^2\ ,
\end{cases}
\end{equation}
so that $I_{\bari}$, $I_{\hati}$ and $I_{i'}$ form orthogonal bases for $\mfrak$, $u(N_c)$ and $u(N_f{-}N_c)$, respectively.
One can associate to $I_i$ vector fields $V_i$ on  U$(N_f)$ and a basis $\{e^i\}=\{e^\bari,e^\hati,e^{i'}\}$ of one-forms 
which is dual to $\{V_i\}$, i.e.~$V_i\lrcorner\, e^j=\de^j_i$.
These one-forms obey the Maurer-Cartan equations
\begin{equation}\label{3.3}
\begin{aligned}
\diff e^\bari &\=-f^{\bari}_{\hatj\bar{k}}\, e^\hatj\wedge e^{\bar k} - f^\bari_{j' \,\bar k}\, e^{j'}\wedge e^{\bar k}\ ,\\
\diff e^\hati &\=-\sfrac12\,f^{\hati}_{\barj\bar{k}}\, e^\barj\wedge e^{\bar k} - \sfrac12\,f^\hati_{\hatj\hat k}\, e^\hatj\wedge e^{\hat k}\ ,\\
\diff e^{i'} &\=-\sfrac12\,f^{i'}_{\barj\bar{k}}\, e^\barj\wedge e^{\bar k} - \sfrac12\,f^{i'}_{j'\, k'}\, e^{j'}\wedge e^{k'}\ ,
\end{aligned}
\end{equation}
where we used the fact that $\Gr(N_c, N_f)$ is a symmetric space.

The Grassmannian $\Mcal = \Gr(N_c, N_f)$ supports an orthonormal frame of one-forms $\{e^\bari\}$ locally giving the U($N_f$)-invariant metric as
\begin{equation}\label{3.6}
\diff s^2_{\Mcal}\=\de_{\bari\barj}\,e^\bari e^\barj \= \de_{\bari\barj}\,e^\bari_\a e^\barj_\b\, \diff X^\a \diff X^\b
\ =:\ g_{\a\b}\,\diff X^\a\diff X^\b  \for \a,\b=1,\ldots,2N_c(N_f{-}N_c),
\end{equation}
where $\{X^\a\}$ is a set of real local coordinates of a point~$X\in\Gr(N_c, N_f)$, 
and $\pa_\a=\pa/\pa X^\a$ will denote derivatives with respect to them.

\medskip

\noindent {\bf Canonical connection.} 
On the principal U($N_c$)-bundle (\ref{2.12}) there exists a unique U($N_f$)-equivariant connection, the so-called canonical connection
(see e.g.~\cite{KNv1}-\cite{HN}),
\begin{equation}\label{3.7}
\Acal_{\Gr}\=\Acal_\a^{\Gr}\,\diff X^\a\=e^\hati I_\hati\=e^\hati_\a I_\hati\,\diff X^\a
\end{equation} 
taking values in $u(N_c)$. It satisfies both Yang--Mills and generalized instanton equations on $\Gr(N_c, N_f)$ \cite{HaTS, HILP, HN}. 
The curvature of the canonical connection (\ref{3.7}) in the bundle (\ref{2.12}) follows as
\begin{equation}\label{3.8}
\Fcal_{\Gr}\=\sfrac12\,\Fcal^{\Gr}_{\a\b}\,\diff X^\a\wedge\diff X^\b
\= -\sfrac12\,f^\hati_{\barj\bar k}\, I_{\hati}\, e^\barj\wedge e^{\bar k} 
\= -\sfrac12\,f^\hati_{\barj\bar k}\, I_{\hati}\, e^\barj_\a e^{\bar k}_\b\; \diff X^\a\!\wedge\diff X^\b\ .
\end{equation} 

\medskip

\noindent {\bf Variation of $\hat\Phi$}. 
By letting $X^\a$ run over $\Mcal$ we obtain a local section $\hat\Phi(X^\a)$ of the bundle (\ref{2.12}).
The infinitesimal changes of this section are given by the covariant derivatives (cf.~\cite{HS, HMS})
\begin{equation}\label{3.9}
\de_\a\hat\Phi\=\pa_\a\hat\Phi + \Acal_\a^{\Gr}\,\hat\Phi\ ,
\end{equation}
where $\Acal_\a^{\Gr}$ are the components of the connection (\ref{3.7}) in the principal U($N_c$) bundle~(\ref{2.12}).

\bigskip

\section{Faddeev model in the infrared limit of 4d YMH}

\noindent {\bf Dependence  on $x^\mu$.} 
Now we return to Yang--Mills--Higgs theory on $\R^{3,1}$. 
In Section~3 we described the moduli space $\Mcal = \Gr(N_c, N_f)$ of vacua for the YMH model (\ref{2.6})-(\ref{2.8}). 
For small exitations around $\Mcal$, in the strong gauge-coupling limit $e^2\gg 1$, the Higgs field $\Phi (x)$ can be considered as a map
\begin{equation}\label{4.1}
\Phi :\quad \R^{3,1}\ \to\ \Gr(N_c, N_f)
\end{equation}
since for $e^2\gg 1$ it should be at a minimum of the Higgs potential (\ref{2.7}). The moduli-space approximation
then postulates that all fields depend on the spacetime coordinates $x=\{x^\mu\}$ only via coordinates $X^\a=X^\a (x)$ on $\Mcal$ 
(see e.g.~\cite{Ma}-\cite{ILP2} and references therein). By substituting $\Phi (X^\a (x) )$ and $\Acal (X^\a (x) )$ into the initial action (\ref{2.6}), 
we obtain an effective field theory describing small fluctuations around the vacuum moduli space $\Mcal$.

\medskip

\noindent {\bf Two-derivative part of effective action}. 
Multiplying (\ref{3.9}) by $\pa_\mu X^\a$, we obtain
\begin{equation}\label{4.2}
\pa_\mu\Phi \= (\pa_\mu X^\a)\,\pa_{\a}\Phi \=(\pa_\mu X^\a)\de_{\a}\Phi-\eps_\mu\Phi 
\ \with\ \eps_\mu = (\pa_\mu X^\a)\Acal_{\a}^{\Gr}\ ,
\end{equation}
where $\eps_\mu\in u(N_c)$  is the pull-back of $\Acal^{\Gr}$ from $\Gr(N_c, N_f)$ to $\R^{3,1}$.
It immediately follows that
\begin{equation}\label{4.3}
D_\mu\Phi \= \pa_\mu\Phi + \Acal_\mu\Phi \= (\pa_\mu X^\a)\de_\a\Phi + (\Acal_\mu - \eps_\mu)\Phi \ .
\end{equation}
We see that $D_\mu\Phi $ are tangent\footnote{
This is a key requirement of the adiabatic approach. It is necessary for the description of {\it small\/} fluctuations around the 
initial moduli space when the dynamical fields are collective coordinates (see e.g.~\cite{HS, HMS, Uhl}).} 
to $C^{\infty} (\R^{3,1}, \Gr(N_c, N_f))$ if 
\begin{equation}\label{4.4}
\Acal_\mu = \eps_\mu\ .
\end{equation}
Substituting  (\ref{4.3})  with  (\ref{4.4}) into the  (\ref{2.6}), we obtain
\begin{equation}\label{4.5}
S_{\textrm{kin}}\=-\int_{\R^{3,1}} \!\!\!\diff^4x\ \eta^{\mu\nu}\ \tr\bigl\{ D_\mu\Phi (D_\nu\Phi)^\+\bigr\}
\=-\sfrac{M^2}{2}\int_{\R^{3,1}} \!\diff^4x\ \eta^{\mu\nu}\,g_{\a\b}\,\pa_\mu X^\a \pa_\nu X^\b \ ,
\end{equation}
where
\begin{equation}\label{4.6}
g_{\a\b}\=\sfrac{2}{M^2}\ \tr\,\bigl\{\de_\a\Phi\,(\de_\b\Phi)^\+\bigr\} \= \de_{\bari\barj}\, e^\bari_\a e^\barj_\b
\end{equation}
are the components of the metric (\ref{3.6}) on $\Gr(N_c, N_f)$ pulled back to $\R^{3,1}$, so $g_{\a\b}(X^\gamma(x))$ now depend on $x$. 
We introduced the mass scale $M$ from (\ref{2.7}) into (\ref{4.6}) to render $g_{\a\b}$ dimensionless. 
Thus, this part of the action~(\ref{2.6}) reduces to the standard non-linear sigma model on $\R^{3,1}$ with the Grassmannian $\Gr(N_c, N_f)$ as its target.

\medskip

\noindent {\bf Four-derivative part of effective action}.  
As discussed earlier, the potential term in (\ref{2.6}) vanishes since $\Phi(x)$ 
takes values in the manifold $\Mcal = \Gr(N_c, N_f)$ of gauge-inequivalent vacua. For calculating the first term in (\ref{2.6}), 
we use (\ref{4.4}), and for the curvature of $\Acal=\Acal_\mu \diff x^\mu$ we obtain
\begin{equation}\label{4.7}
\Fcal \=\diff \Acal +  \Acal\wedge \Acal  \= \sfrac12\,\Fcal_{\mu\nu}\,\diff x^\mu\wedge\diff x^\nu \=
-\sfrac1{2}\,f^\hati_{\barj\bar k}\, I_\hati\, e^\barj_\a e^{\bar k}_\b\, \pa_\mu X^\a  \pa_\nu X^\b\,
\diff x^\mu\wedge\diff x^\nu \  ,
\end{equation}
allowing one to extract the components $\Fcal_{\mu\nu}$. Substituting (\ref{4.7}) into (\ref{2.6}) we arrive at
\begin{equation}\label{4.8}
\begin{aligned}
S_{\textrm{Fad}} &\= -\sfrac{1}{2e^2 }\int_{\R^{3,1}} \!\!\!\diff^4x\ \tr\bigl(\Fcal_{\mu\nu}^\+\Fcal^{\mu\nu}\bigr) \\
&\= -\sfrac{1}{4 e^2} \int_{\R^{3,1}} \!\!\!\diff^4x\ \de_{\hati\hatj} f^\hati_{\bar l\bar k}f^\hatj_{\bar m\bar n}\, 
e^{\bar l}_\a  e^{\bar k}_\b e^{\bar m}_\gamma e^{\bar n}_\de \,\pa_\mu X^\a \pa_\nu X^\b\pa^\mu X^\gamma\pa^\nu X^\de\ ,
\end{aligned}
\end{equation}
where $\pa^\mu:=\eta^{\mu\sigma}\pa_\sigma$. 
Thus, in the infrared limit the Yang--Mills--Higgs action (\ref{2.6}) is reduced to the Faddeev action,
\begin{equation}\label{4.9}
S_{\textrm{eff}} \= -\int_{\R^{3,1}}\!\!\!\diff^4x\ \Bigl\{\frac{M^2}{2} g_{\a\b}\pa_\mu X^\a \pa^\mu X^\b\ +\
\frac{1}{4 e^2}\de_{\hati\hatj} f^\hati_{\bar l\bar k}f^\hatj_{\bar m\bar n}\,e^{\bar l}_\a  e^{\bar k}_\b e^{\bar m}_\gamma e^{\bar n}_\de\,
\pa_\mu X^\a \pa_\nu X^\b\pa^\mu X^\gamma\pa^\nu X^\de\Bigr\}
\end{equation}
for scalar fields $X^\a$ with values in the Grassmannian $\Gr(N_c, N_f)$.

\bigskip
%\newpage

\section{A$_2{\oplus}$A$_2$-quiver gauge theory}

\noindent {\bf Fields.}  
It is possible to obtain not only the Faddeev model but also the standard Skyrme model from Yang--Mills--Higgs theory in four dimensions.
To achieve this, we should consider a 4d YMH model with a group manifold, say U($N$), as the moduli space $\Mcal$ of vacua. 
The simplest way to do this is to specialize the model (\ref{2.6}) to $N_f=2N_c=:2N$ but with a potential different from~(\ref{2.7}). 
We parametrize
\begin{equation}
\Phi =: (\phi_- , \phi_+) \ \with\ \phi_\pm \in\textrm{Mat}(N,N;\Cbb)\ .
\end{equation}
Thus, we have a $u(N)$-valued gauge field $\Fcal$, an $N{\times}2N$ complex Higgs field $\Phi=(\phi_-,\phi_+)$, 
the group of gauge transformations $\Gcal=C^\infty (\R^{3,1},$ U$(N))$ and transformations  (\ref{2.2})-(\ref{2.5}) for $N_f=2N_c=2N$.

\medskip

\noindent {\bf Action.} 
We consider the Yang--Mills--Higgs (YMH) action functional 
\begin{equation}\label{5.1}
S\=-\int_{\R^{3,1}} \!\!\!\!\diff^4x\ \bigl\{
\tr\,\bigl(\sfrac{1}{2e^2}\,\Fcal^\+_{\mu\nu}\Fcal^{\mu\nu} +  D_{\mu}\Phi (D^{\mu}\Phi)^\+\bigr) +\sfrac{e^2}{4}\, V(\Phi )\bigr\}\ ,
\end{equation}
and the two-term potential
\begin{equation}\label{5.2}
V(\Phi )\=\tr\,\bigl(m^2\unity_{N} - \phi_-\phi_-^\+\bigr)^2\ +\ \tr\,\bigl(m^2\unity_{N} - \phi_+\phi_+^\+\bigr)^2
\end{equation}
with a mass parameter $m$. This action can be obtained from A$^+_2{\oplus}$A$^-_2$ quiver gauge theory 
(see e.g.~\cite{PS, LPS} and references therein) corresponding  to a direct sum of quivers
\begin{equation}\label{5.3}
A_2^\pm : \quad \Cbb^N\ \stackrel{\phi_\pm}{\longrightarrow}\ \Cbb^{N}\ ,
\end{equation}
where four copies of $\Cbb^N$ at four vertices carry the fundamental U($N$) representation, 
and the arrows $\phi_\pm$ denote maps between them.

The form (\ref{5.2}) of the Higgs potential breaks the flavour group U$(2N)$ to the subgroup $G=\textrm{U}_-(N)\times\textrm{U}_+(N)$. 
Let $\{I_i\}$ be a basis of the Lie algebra $\gfrak=\textrm{Lie}G=u_-(N)\oplus u_+(N)$ realized as $2N{\times}2N$ block-diagonal matrices 
with the normalization $\tr(I_i I_j)=-\sfrac12\de_{ij}$ for $i=1,\ldots,2N^2$. The covariant derivative in (\ref{5.1}) reads
\begin{equation}\label{5.4}
D_\mu\Phi \= (D_\mu\phi_-\,,\, D_\mu\phi_+) \ \with\  D_\mu\phi_\pm =\pa_\mu\phi_\pm  +\Acal_\mu\phi_\pm\ ,
\end{equation}
with a $u(N)$-valued gauge potential $\Acal = \Acal_\mu\diff x^\mu$.

\medskip

\noindent {\bf Vacua}. 
The energy density of YMH configurations described by the action (\ref{5.1}) 
has the form (\ref{2.8}) with $V(\Phi )$ given by (\ref{5.2}).  The vacuum configurations are defined by (\ref{2.9}), which implies
\begin{equation}\label{5.5} 
\hat\phi_-\hat\phi_-^\+ \= m^2\,\unity_N\und \hat\phi_+\hat\phi_+^\+ \= m^2\,\unity_N\  .
\end{equation}
Equations (\ref{5.5}) are solved by some
\begin{equation}\label{5.6} 
(\hat\phi_- , \hat\phi_+)\in {\rm U_-}(N){\times}{\rm U}_+(N)=\widetilde\Mcal
\end{equation}
subject to global gauge transformations 
\begin{equation}\label{5.7} 
(\hat\phi_- , \hat\phi_+)\ \mapsto\ (h\hat\phi_- , h\hat\phi_+)\ \for  h\in {\rm U}(N)\ .
\end{equation}
The group U$(N)$ acts freely on the vacuum manifold $\widetilde\Mcal$ by left multiplication, and one can define the projection
\begin{equation}\label{5.8}
\pi\ :\quad \widetilde\Mcal\ \stackrel{{\rm U}(N)}{\longrightarrow}\ \Mcal \qquad\textrm{via}\qquad
(\hat\phi_- , \hat\phi_+)\ \longmapsto\ (\unity_N,\hat\phi) \ \with \hat\phi =\hat\phi_-^{-1} \hat\phi_+\ .
\end{equation}
Hence, the moduli space of vacua 
\begin{equation}\label{5.10}
\Mcal \= {\rm U}(N) \backslash [{\rm U}_-(N)\times  {\rm U}_+(N)]
\end{equation}
is diffeomorphic to the group manifold U($N$), any element of which can be obtained from a reference vacuum $\hat\Phi_0$. 
We choose $\sfrac1m\,\hat\Phi_0=(\unity_N,\unity_N)$ so that the isotropy group for the right $G$~action is
\begin{equation}\label{5.11}
\textrm{U}(N) \= \bigl\{g\in G :\ \hat\Phi_0\,g= h\,\hat\Phi_0\quad\textrm{for some}\ h\in {\rm U}(N)\bigr\}\ .
\end{equation}
It is obvious that $g=$diag$(h,h)$, i.e.~the isotropy group is 
\begin{equation}
\textrm{diag}(G)\ \cong\ \textrm{U}_\textrm{diag}(N) \= \textrm{U}(N)\ =:\ H\ ,
\end{equation}
and the global gauge transformations form the stability subgroup in a realization of the group manifold U$(N)$ 
as the coset space $H\backslash G$ in (\ref{5.10}).
This is also seen from the fact that $(h\hat\phi_-)^{-1}(h\hat\phi_+)=\hat\phi_-^{-1}\hat\phi_+$, 
i.e.~$\hat\phi$ is inert under the action of $H$. From (\ref{5.8}) it follows the decomposition
\begin{equation}\label{5.12}
\gfrak \=u_-(N)\oplus u_+(N)\=\mfrak\oplus\hfrak\=\mfrak\oplus u(N)_\textrm{diag}
\ \with \hfrak = \bigl\{(\eta , \eta)\,\bigm|\, \eta\in u(N)\bigr\}\ .
\end{equation}

\medskip

\noindent {\bf Geometry of $H\backslash G$}.  
The geometry of a group manifold considered as a homogeneous space has some characteristic features 
(see e.g.~\cite{KNv2, KMR, Ha2}) which we briefly describe here. In the split (\ref{5.12}), 
$\mfrak$ is not necessarily orthogonal to $\hfrak$ with respect to the Cartan--Killing form. 
In fact, there are three natural reductive decompositions of $\gfrak$ with the following versions of~$\mfrak$:
\begin{equation}\label{5.13}
\mfrak_0 = \bigl\{(-\th, \th)\bigr\}\ ,\qquad
\mfrak_- = \bigl\{(-\th ,0)\bigr\}\ ,\qquad
\mfrak_+ = \bigl\{(0, \th)\bigr\}\ ,\quad\with \th\in u(N)\ .
\end{equation}
The first case yields $H\backslash G$ as a symmetric space with $\mfrak_0$ orthogonal to $\hfrak$. 
With the choice $\mfrak_+$ or $\mfrak_-$ the coset (\ref{5.10}) 
becomes a nonsymmetric homogeneous manifold. Obviously, $\mfrak\cong u(N)$ in all three cases.
The choices of $\mfrak_0$, $\mfrak_-$ and $\mfrak_+$ correspond to the gauges 
$\hat\phi_-{=}\,\hat\phi_+^{\+}$, $\hat\phi_+{=}\,m\unity_N$ and $\hat\phi_-\,{=}\,m\unity_N$, respectively, 
which determine different coset representatives, 
i.e.~sections of the bundle (\ref{5.8}) with $\widetilde\Mcal = G$ and $\Mcal =H\backslash G$.

We split the basis of $\gfrak$ according to the decomposition (\ref{5.12}),
\begin{equation}
\{I_i\} \= \{I_\bari,I_\hati\} \ \with\ \begin{cases}
\ \bari \!\!&=\ 1,\ldots,N^2 \qquad\quad\for \mfrak \ ,\\
\ \hati \!\!&=\ N^2{+}1,\ldots,2N^2 \ \for \hfrak\ .
\end{cases}
\end{equation}
We have an orthonormal frame of one-forms $\{e^\bari\}$ on $H\backslash G$, 
the metric (\ref{3.6}) with $\a,\b=1,\ldots,N^2$ and the canonical connection 
$\Acal^\textrm{can}=e^\hati \,I_\hati = e^\hati_\a \,I_\hati\,\diff X^\a$ 
for all three cases $\mfrak_0$, $\mfrak_-$ and $\mfrak_+$. However, the Maurer--Cartan equations depend on the case:
\begin{equation}\label{5.14}
\begin{aligned}
\mfrak_0:\quad\diff e^\bari =-f^{\bari}_{\hatj\bar{k}}\, e^\hatj\wedge e^{\bar k} 
\und
\diff e^\hati = -\sfrac12\,f^\hati_{\barj\bar k}\, e^\barj\wedge e^{\bar k}-\sfrac12\,f^{\hati}_{\hatj\hat{k}}\, e^\hatj\wedge e^{\hat k}\ ,\\
\mfrak_-:\quad\diff e^\bari =-f^{\bari}_{\hatj\bar{k}}\, e^\hatj\wedge e^{\bar k} + \sfrac12\,f^\bari_{\barj\bar k}\, e^\barj\wedge e^{\bar k}
\und
\diff e^\hati =-\sfrac12\,f^{\hati}_{\hatj\hat{k}}\, e^\hatj\wedge e^{\hat k} \ ,\\
\mfrak_+:\quad\diff e^\bari =-f^{\bari}_{\hatj\bar{k}}\, e^\hatj\wedge e^{\bar k} - \sfrac12\,f^\bari_{\barj\bar k}\, e^\barj\wedge e^{\bar k}
\und
\diff e^\hati =-\sfrac12\,f^{\hati}_{\hatj\hat{k}}\, e^\hatj\wedge e^{\hat k}\ .
\end{aligned}
\end{equation}
Furthermore, on the group manifold  (\ref{5.10}) one can introduce a family of connections
\begin{equation}\label{5.17}
\Acal^\vk_{{\rm U}(N)} \= \vk\, e^\hati I_\hati \= \vk\, e^\hati_\a I_\hati \diff X^\a \ =:\ \Acal^\vk_{\a}\,\diff X^\a \ \with \vk\in\R
\end{equation}
with curvature 
\begin{equation}\label{5.18}
\Fcal^\vk_{{\rm U}(N)}\=\sfrac12\,\vk (\vk{-}1)\,f^{\hati}_{\hatj\hat{k}}\, I_{\hati}\,e^\hatj\wedge e^{\hat k}
-\sfrac12\,\vk \,f^\hati_{\barj\bar k}\, I_{\hati}\, e^\barj\wedge e^{\bar k}\ .
\end{equation}
For the cases $\mfrak_\pm$ the last term in  (\ref{5.18}) vanishes. The connection (\ref{5.17}) is the unique $G$-equivariant 
family of connections on the bundle~(\ref{5.8})~\cite{KNv2, KMR}.

\medskip

\noindent {\bf Variation of $\hat\Phi$}. 
In the following we adopt the gauge $\sfrac1m\,\hat\phi_-=\unity_N$ fixing $\mfrak=\mfrak_+$, so $f^\hati_{\barj\bar k}=0$ in (\ref{5.18}).
Then, abbreviating $\hat\phi_+\equiv\hat\phi$,
\begin{equation}\label{5.19}
\Acal^\vk_{{\rm U}(N)}\=\vk\,\hat\phi\,(\pa_\a\hat\phi^{-1})\,\diff X^\a\quad\Rightarrow\quad  
\Acal^\vk_{\a} \=\vk\,\hat\phi\,\pa_\a \hat\phi^{-1}\ ,
\end{equation}
and letting $X^\a$ run over $\Mcal\cong\textrm{U}(N)$ we obtain a local section $\hat\Phi(X^\a)=m(\unity_N,\hat\phi(X^\a))$ of the bundle (\ref{5.8}). 
Infinitesimal changes of this section are given by the covariant derivatives (cf.~\cite{HS, HMS}) 
\begin{equation}\label{5.20}
\de_\a \hat\Phi \= \pa_\a \hat\Phi +\Acal^\vk_{\a}\hat\Phi \= m \bigl(\Acal^\vk_{\a}\, ,\, \pa_\a \hat\phi +\Acal^\vk_{\a}\hat\phi \bigr)\ .
\end{equation}

\medskip

\section{Skyrme model in the infrared limit of 4d YMH theory}

\noindent The derivation of the Skyrme model as an effective theory for the 4d YMH model (\ref{5.1}) is similar to the derivation 
of the Faddeev model from the YMH action  (\ref{2.6}). The main difference is that now the vacuum moduli space $\Mcal=H\backslash G=\textrm{U}(N)$
is a group manifold, whose geometry was described in Section~5.  According to the philosophy of the adiabatic method, we assume that the gauge potential 
$\Acal = \Acal_\mu\diff x^\mu$ and the Higgs field $\Phi$ depend on the $\R^{3,1}$ coordinates $x$ only via real coordinates $X^\a =X^\a(x)$ on 
U($N$), and we substitute $\Acal (X^\a(x))$ and $\Phi (X^\a(x))$ into the action  (\ref{5.1}) by using results of Section~5.

\medskip

\noindent {\bf Kinetic term}. Multiplying  (\ref{5.20}) by $\pa_\mu X^\a$, we obtain 
\begin{equation}\label{6.1}
D_\mu\Phi \=\pa_\mu X^\a \ \de_\a\Phi \ +\ (A_\mu - \eps_\mu )\Phi  \ ,
\end{equation}
where $ \eps_\mu = (\pa_\mu X^\a)\Acal_\a^\vk\in u(N)$ is the pull-back of $\Acal^\vk_{{\rm U}(N)}$ from U$(N)$ to $\R^{3,1}$. 
To render $(D_\mu\Phi )\Phi^{\+}$ tangent to $C^\infty (\R^{3,1}, {\rm U}(N))$, we choose 
\begin{equation}\label{6.2}
\Acal_\mu \=\eps_\mu \= \vk\,(\pa_\mu X^\a)\,\phi\,\pa_\a \phi^{-1} \= \vk\,\phi\,\pa_\mu \phi^{-1}\ ,
\end{equation}
where $\phi$ is a U($N$)-valued function. Notice that (\ref{5.19}) and (\ref{5.20}) imply 
\begin{equation}\label{6.3}
\de_\a\Phi \= -m\bigl(\vk\,(\pa_\a\phi)\, \phi^\+ , \,(\vk{-}1)\,\pa_\a\phi\bigr) \qquad\Rightarrow\qquad
D_\mu\Phi \= -m\bigl(\vk\,(\pa_\mu\phi)\, \phi^\+ , \,(\vk{-}1)\,\pa_\mu\phi\bigr) \ .
\end{equation}
Substituting (\ref{6.1})--(\ref{6.3}) into  (\ref{5.1}), we obtain
\begin{equation}\label{6.4}
S_{\textrm{kin}}\=-\int_{\R^{3,1}} \!\!\!\diff^4x\ \eta^{\mu\nu}\ \tr\bigl\{ D_\mu\Phi (D_\nu\Phi)^\+\bigr\}
\=\sfrac14\,f^2_\pi\int_{\R^{3,1}} \!\!\!\diff^4x\ \eta^{\mu\nu}\ \tr (R_\mu R_\nu )
\end{equation}
with 
\begin{equation}\label{6.5}
R_\mu:=\phi\,\pa_\mu\phi^{-1} \und
\sfrac14\,f^2_\pi =\bigl(\vk^2+ (\vk{-}1)^2\bigr)\,m^2\ ,
\end{equation}
where $f_\pi$ may be interpreted as the pion decay constant.
Thus, this part of the action (\ref{5.1}) reduces to the standard non-linear sigma model on $\R^{3,1}$
with a U$(N)$ target space.

\medskip

\noindent {\bf Skyrme term}. 
For calculating the $\Fcal^2$-terms in (\ref{5.1}) we employ (\ref{6.2}) and find
\begin{equation}\label{6.6}
\Fcal\=\diff\Acal + \Acal\wedge\Acal\= \vk(\vk{-}1)\; \phi\,\diff\phi^{-1}\wedge \phi\,\diff\phi^{-1}
\=\sfrac1{2}\, \vk(\vk{-}1)\;[R_\mu , R_\nu]\,\diff x^\mu\wedge\diff x^\nu
\end{equation}
since $\Acal_\mu=\vk\,\phi\,\pa_\mu\phi^{-1}=\vk R_\mu$ after the pull-back to $\R^{3,1}$. Substituting  (\ref{6.6}) into  (\ref{5.1}), we obtain
\begin{equation}\label{6.7}
S_{\textrm{Sky}} \= 
-\frac{1}{2e^2}\,\int_{\R^{3,1}} \!\!\diff^4x\ \tr\bigl(\Fcal_{\mu\nu}^\+\Fcal^{\mu\nu}\bigr) \=
\frac{1}{32\zeta^2} \int_{\R^{3,1}}\!\!\diff^4x\ \eta^{\mu\lambda}\eta^{\nu\sigma}\,\tr\bigl([R_\mu, R_\nu][R_\lambda, R_\sigma]\bigr) \ ,
\end{equation}
where 
\begin{equation}\label{6.8}
\frac{1}{32\zeta^2}\=\frac{\vk^2(\vk{-}1)^2}{8e^2}\ ,
\end{equation}
and $\zeta$ is the dimensionless Skyrme parameter. Hence, in the infrared limit 
the Yang--Mills--Higgs action (\ref{5.1}) is reduced to the action of the Skyrme model,
\begin{equation}\label{6.9}
S_{\textrm{eff}} \= \int_{\R^{3,1}}\!\!\diff^4x\,\Bigl\{\frac{f^2_\pi}{4}\,\eta^{\mu\nu}\, \tr\bigl(R_\mu R_\nu\bigr)\ +\
\frac{1}{32\zeta^2}\,\eta^{\mu\lambda}\eta^{\nu\sigma}\,\tr\bigl([R_\mu, R_\nu][R_\lambda, R_\sigma]\bigr)\Bigr\}\ .
\end{equation}
Thus, both Skyrme and Faddeev models appear as effective field theories in the infrared of Yang--Mills--Higgs models.

\bigskip

\noindent {\bf Acknowledgements}

\noindent
This work was partially supported by the Deutsche Forschungsgemeinschaft grant LE 838/13.
It is based upon work from COST Action MP1405 QSPACE, supported by COST (European Cooperation in Science and Technology).

\newpage
%\bigskip

\end{document}